\documentclass[journal = ancac3]{achemso}
\setkeys{acs}{etalmode=truncate,maxauthors=10}

\usepackage{amsmath}
\usepackage{amscd}
\usepackage{amssymb}
\usepackage{amsthm}
\usepackage{amsfonts}
\usepackage{amstext}
\usepackage{textcomp}
\usepackage{graphicx}
\usepackage{dcolumn}
\usepackage{bm}
\usepackage[usenames]{color}
\usepackage[normalem]{ulem}
\usepackage{epsfig} 
\usepackage{epstopdf}
\usepackage{natbib}
\usepackage{subfig}
\usepackage{soul}
\usepackage{lineno}
\usepackage{atbegshi}% http://ctan.org/pkg/atbegshi
\AtBeginDocument{\AtBeginShipoutNext{\AtBeginShipoutDiscard}}
\setulcolor{red}

\newcommand{\onlinecite}[1]{\hspace{-1 ex} \nocite{#1}\citenum{#1}}

\usepackage{natbib}
\definecolor{klimgreen}{rgb}{0.0, 0.7, 0.1}

\title{Interfacing Quantum Spin Hall and Quantum Anomalous Hall insulators: Bi bilayer on MnBi$_2$Te$_4$-family materials}

\author{I.\,I.~Klimovskikh}
\email{ilya.klimovskikh@dipc.org}
\affiliation{Donostia International Physics Center (DIPC), 20018 Donostia-San Sebasti\'{a}n, Basque Country, Spain}

\author{S.\,V.~Eremeev}
\affiliation{Institute of Strength Physics and Materials Science,
Russian Academy of Sciences, 634055 Tomsk, Russia}

\author{D.\,A.~Estyunin}
\affiliation{Saint Petersburg State University, 198504 Saint Petersburg, Russia}

\author{S.\,O.~Filnov}
\affiliation{Saint Petersburg State University, 198504 Saint Petersburg, Russia}

\author{K. Shimada}
\affiliation{Hiroshima Synchrotron Radiation Center, Hiroshima University, Hiroshima, Japan}

\author{V.~A.~Golyashov}
 \affiliation{Synchrotron Radiation Facility SKIF, Boreskov Institute of Catalysis, Siberian Branch, Russian Academy of Sciences, Kol’tsovo 630559, Russia}
\alsoaffiliation{Saint Petersburg State University, 198504 Saint Petersburg, Russia}

\author{O.~E.~Tereshchenko}
 \affiliation{Synchrotron Radiation Facility SKIF, Boreskov Institute of Catalysis, Siberian Branch, Russian Academy of Sciences, Kol’tsovo 630559, Russia}
\alsoaffiliation{Saint Petersburg State University, 198504 Saint Petersburg, Russia}

   \author{K.~A.~Kokh}
\affiliation{
Sobolev Institute of Geology and Mineralogy, Siberian Branch, Russian Academy of Sciences, 630090 Novosibirsk, Russia}
\alsoaffiliation{Saint Petersburg State University, 198504 Saint Petersburg, Russia}

\author{A.~S.~Frolov}
\affiliation{Lomonosov Moscow State University, Leninskie Gory 1/3, Moscow 119991, Russia}

\author{A.~I.~Sergeev}
\affiliation{Lomonosov Moscow State University, Leninskie Gory 1/3, Moscow 119991, Russia}

\author{V.~S.~Stolyarov}
\affiliation{Donostia International Physics Center (DIPC), 20018 Donostia-San Sebasti\'{a}n, Basque Country, Spain}

\author{V.~Mikšić Trontl}
\affiliation{Centre for Advanced Laser Techniques, Institute of Physics, 10000 Zagreb, Croatia}

\author{L.\ Petaccia}
\affiliation{Elettra Sincrotrone Trieste, Strada Statale 14 km 163.5, 34149 Trieste, Italy}

\author{G.\ Di Santo}
\affiliation{Elettra Sincrotrone Trieste, Strada Statale 14 km 163.5, 34149 Trieste, Italy}

\author{M.\ Tallarida}
\affiliation{ALBA Synchrotron Light Source, Cerdanyola del Vallès, 08290 Barcelona, Spain}

\author{J.\ Dai}
\affiliation{ALBA Synchrotron Light Source, Cerdanyola del Vallès, 08290 Barcelona, Spain}

\author{S.\ Blanco-Canosa}
\affiliation{Donostia International Physics Center (DIPC), 20018 Donostia-San Sebasti\'{a}n, Basque Country, Spain}
\alsoaffiliation{IKERBASQUE, Basque Foundation for Science, 48013 Bilbao, Spain}

\author{T.\ Valla}
\affiliation{Donostia International Physics Center (DIPC), 20018 Donostia-San Sebasti\'{a}n, Basque Country, Spain}

\author{A.\,M.~Shikin}
\affiliation{Saint Petersburg State University, 198504 Saint Petersburg, Russia}

\author{E.\,V. Chulkov}

\affiliation{Donostia International Physics Center (DIPC), 20018 Donostia-San Sebasti\'{a}n, Basque Country, Spain}
\alsoaffiliation{Saint Petersburg State University, 198504 Saint Petersburg, Russia}
\alsoaffiliation{Departamento de Polímeros y Materiales Avanzados: Física, Química y Tecnología,
Facultad de Ciencias Qumicas, Universidad del País Vasco UPV/EHU, 20080 San
Sebastián/Donostia, Basque Country, Spain}

\begin{document}
\maketitle

\newpage
\begin{abstract}
Meeting of non-trivial topology with magnetism results in novel phases of matter, such as Quantum Anomalous Hall (QAH) or axion insulator phases.  Even more exotic states with high and tunable Chern numbers are expected at the contact of intrinsic magnetic topological insulators (IMTIs) and 2D topological insulators (TIs).Here we synthesize a heterostructures composed of 2D TI and 3D IMTIs, specifically of bismuth bilayer on top of MnBi$_2$Te$_4$-family of compounds and study their electronic properties by means of angle-resolved photoelectron spectroscopy (ARPES) and density functional theory (DFT). The epitaxial interface is characterized by hybridized Bi and IMTI electronic states. The  Bi bilayer-derived states on  different members of MnBi$_2$Te$_4$-family of materials are similar, except in the region of mixing with the topological surface states of the substrate. In that region, the new, substrate dependent interface Dirac state is observed. Our \emph{ab initio} calculations show rich interface phases with emergence of exchange split 1D edge states, making the Bi/IMTI heterostructures promising playground for observation of novel members in the family of quantum Hall effects.
\end{abstract}

\section*{Introduction}

Discovery of intrinsic magnetic topological insulators (IMTIs)  boosts the research on quantum anomalous Hall (QAH) effect and axion electrodynamics and give a new hope for Majorana zero modes observation.\cite{Hirahara, Hirahara2020, Otrokov_2019_Nature, Zhang_2019, Li_2019, Gong_2019, Lee_2019, Aliev_2019, Hao_2019, Chen_2019, Swatek_2020, Estyunin_2020, Gui2019} The QAH phase is characterzed by non-zero Chern number ($C=1$), 
topologically protected chiral edge states and quantized Hall conductance at zero magnetic field.\cite{chang2023} The most studied IMTI compound, MnBi$_2$Te$_4$ (MBT), is a layered van-der-Waals crystal.\cite{Otrokov_2019_Nature, Klimovskikh_2020, Rienks2019} Predicted antiferromagnetic interlayer ordering had been confirmed experimentally, as well as the exchange interaction impact on the topological band structure, fulfilling the requirements for QAH effect observation.\cite{Deng_2020} 

Apart from the QAH with $C=1$, it is possible to realize the topological materials with higher and tunable Chern numbers  whose multiple dissipationless edge conduction channels could significantly improve the performance of quantum devices.\cite{Wang2013, Bosnar2023} Recently such a tunable high-Chern number phase has been predicted for the interface of 2D topological insulators and MnBi$_2$Te$_4$. \cite{Xue_arxiv_2212_12905, li2024multiplemode} As an exciting example in Ref.~[\onlinecite{Xue_arxiv_2212_12905}] authors theoretically demonstrate the  multiple topological edge states and switchable Chern number ($C=\pm1, \pm3$) for the contact of bismuth bilayer with a single septuple layer of MBT. 
 
 Ultrathin bismuth films represent one of the most promising 2D topological materials, whose unique properties had been successfully demonstrated in various structures, such as bismuthene and Bi(110) or Bi(111) films.\cite{Shen_2019, Yeom_PRB2016, https://doi.org/10.1002/pssr.202000131, Reis2017, ACS2016, NanoLett2017, doi:10.1126/sciadv.aba2773} Due to the large spin-orbit interaction, the band gap in these 2D materials is inverted, resulting in Quantum Spin Hall (QSH) phase and 1D spin-polarized states emerge at the edges of bismuth layer islands.\cite{NanoLett2017, PhysRevB.98.245108, Drozdov2014, Kim_PRB2014}  Experimentally, the most accessible form is Bi(111) bilayer. Its topological properties strongly depend on the interaction with the  substrate.\cite{https://doi.org/10.1002/pssr.202000131, PhysRevB.98.245108,Drozdov2014} Due to a good lattice mach, materials like Bi$_2$Te$_3$ and other 3D TIs are generally suitable substrates for Bi bilayer films. Interestingly, the interaction of the  topological surface states of the substrate with Bi-related bands does not destroy the QSH phase and 1D-edge channels, as demonstrated in Refs.~[\onlinecite{Kim_PRB2014, PhysRevB.90.235401, PhysRevLett.107.166801}].  
 Depending on the interaction with the substrate, the contact of Bi bilayer with MBT may also show other, yet undetected types of quantum Hall effect, such as time-reversal symmetry broken QSH.\cite{PhysRevB.91.041303, SciRep2015, PhysRevLett.107.066602, PhysRevB.86.035104, Naumov2023}

 In this study we synthesize the heterostructures consisting of Bi-bilayer and magnetic topological insulators of  the MnBi$_2$Te$_4$ family. By means of  photoelectron spectroscopy (PES)  and Low Energy Electron Diffraction (LEED) the epitaxial growth mode is confirmed, and electronic structure of the interfaces is studied by means of ARPES. Although the interaction between Bi-BL and IMTI surfaces is relatively weak,  the topological surface  states  mix with the Bi-BL states, resulting in a specific interface electronic structure, supported by DFT calculations. Resulting band structure of the Bi-BL exhibits distinctive set of hole-like states, which weakly depend on the substrate composition. On the other hand, the interface Dirac cone-like feature around the $\bar \Gamma$ point  is formed only for IMTI substrates with well-pronounced topological surface states within the bulk band gap. Furthermore, \emph{ab initio} calculations predict the emergence of the 1D  topological states at the edges of Bi-BL, that are exchange split due to interaction with the magnetic substrate.

\section*{Results}

\subsection*{Bi bilayer on MnBi$_2$Te$_4$}

Bulk MnBi$_2$Te$_4$ material consists of septuple layer blocks Te-Bi-Te-Mn-Te-Bi-Te, that stack together forming van der Waals (vdW) crystal. Therefore, cleavage of bulk MnBi$_2$Te$_4$ crystal in ultra-high vacuum (UHV) results in formation of atomically clean Te-terminated (0001) surface. In-plane lattice constant of 4.33  \AA~ in MnBi$_2$Te$_4$ is almost the same as  in Bi$_2$Te$_3$ (4.38 \AA) and close to the one in Bi(111) (4.54 \AA). Hence, it can be expected that  deposition of Bi on top of MnBi$_2$Te$_4$ would lead to epitaxial growth of Bi(111) overlayers, similarly to Bi-bilayer on Bi$_2$Te$_3$ case.\cite{PhysRevLett.107.166801, doi10.1073/pnas.1218104110, PhysRevB.89.155116} Therefore, in order to fabricate and characterize the Bi bilayer on MBT we have deposited the corresponding amount of Bi on freshly cleaved MBT surface, slightly anneal it and carried out  LEED and PES measurements to test the structural properties of the film and then performed the ARPES studies.

\begin{figure*}
\includegraphics[width=\linewidth]{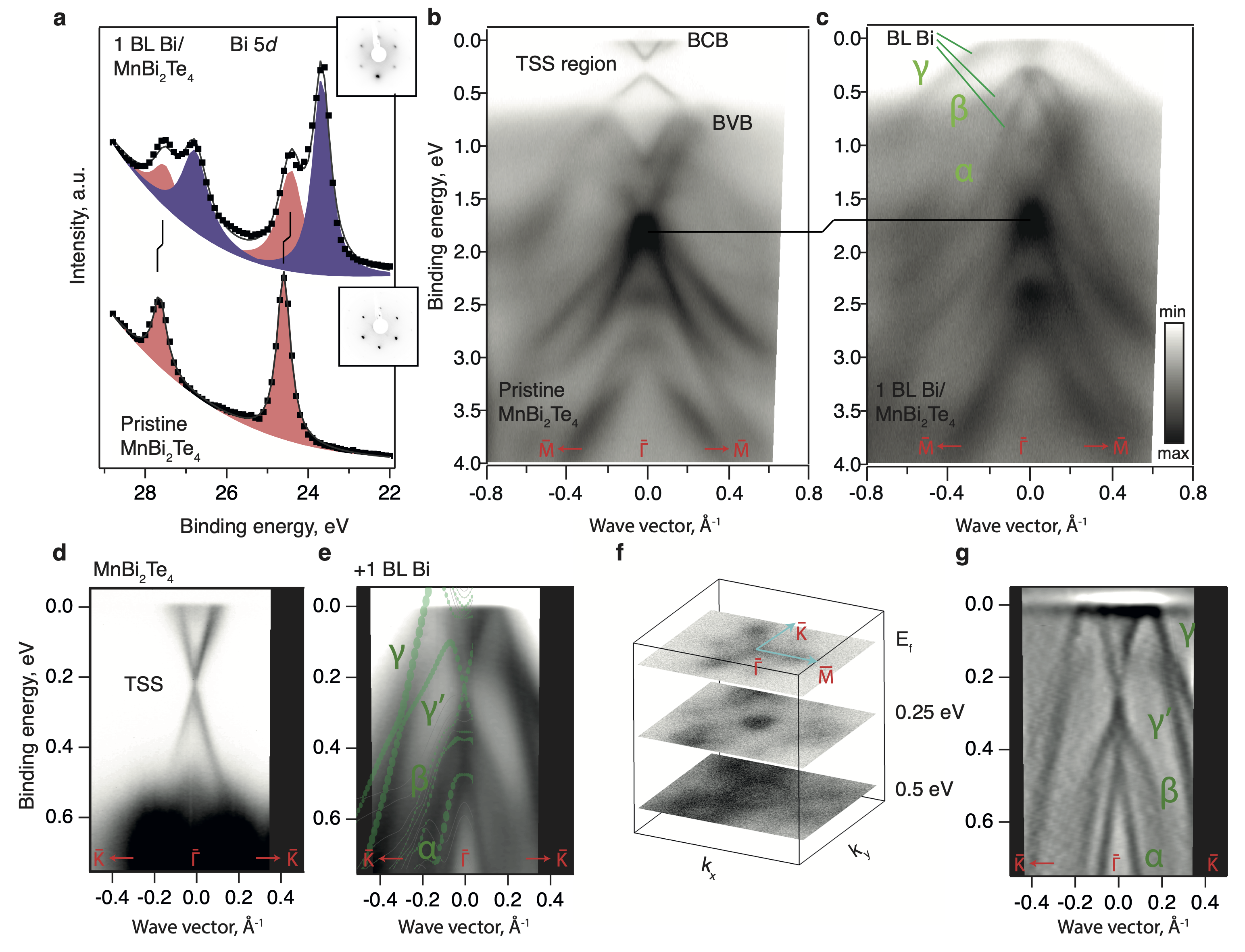}
\caption{
(a) Experimental core level spectra and fitting data of Bi-5$d$  lines for MnBi$_2$Te$_4$ before (lower spectrum) and after (upper spectrum) Bi-bilayer deposition. Data were taken at a photon energy of 40 eV. Upper and lower insets show corresponding LEED images taken at the primary electron energy of 80 eV. (b,c) ARPES dispersion relations in the wide energy region in the $\bar{\Gamma}\bar{\mathrm{M}}$ direction of BZ of MnBi$_2$Te$_4$  taken at temperature of 76 K and at a photon energy of 21.2 eV before (b) and after (c) Bi deposition. (d,e) ARPES dispersion relations in the low energy region in the $\bar{\mathrm{K}}-\bar{\Gamma}-\bar{\mathrm{K}}$ direction of BZ of MnBi$_2$Te$_4$  taken at  $17~\,\mathrm{K}$ at a photon energy of 28 eV  before (d)  and after (e) Bi deposition. DFT calculated spectrum (green) is imposed on the left side of panel (e). (f) Constant energy maps in the $k_\|$-space taken at various binding energies. (g) Curvature plot (the procedure is described in Ref.[\cite{Curvature}] of the ARPES data at (e).} 
 \label{fig1}
\end{figure*}

 PES spectra of Bi 5$d$ lines for MBT surface before  and after  bismuth deposition are presented in Fig.~\ref{fig1}(a). Binding energies (BEs) of Bi 5$d_{5/2,~3/2}$ peaks are 24.6 and 27.7 eV respectively, in accordance with previous works.\cite{Li_2020PCCP, Makarova_2021, PhysRevB.106.155305, Klim2022} Deposition of Bi atoms results in a small shift of Bi 4$d$ line by 0.2 eV towards the lower binding energies.  Besides the slight change of the  main peaks positions Bi 5$d$ spectrum exhibits the appearance of additional peaks (violet color), with the BEs of  23.7  and 26.8 eV. The binding energy of these additional components corresponds to one Bi-bilayer, as shown in Refs.~[\onlinecite{PhysRevB.86.241101, PhysRevB.88.081108, klim2017}]. Taking into account the exponential decay of the photoelectron signal in depth the effective thickness of the Bi film can be estimated as $d= \lambda *ln(1+2/7*I_{\mathrm{Bi}_2}/I_{\mathrm{MBT}})$, where $\lambda$ is the mean free path of the photoelectron with a given kinetic energy  and $I_{\mathrm{Bi}_2}/I_{\mathrm{MBT}}$ is ratio of the intensities of  Bi-BL and substrate contributions. Using this formula we obtain the effective thickness of $\approx$ 4 \AA, that corresponds to a coverage of one bismuth bilayer (Bi-BL height is 3.8 \AA). LEED data clearly demonstrate the $1\times1$ structure of the Bi overlayer on MBT, similar as in case of Bi$_2$Te$_3$ substrate.

In order to have a look into the band structure of the synthesized interface we have measured angle-resolved photoelectron spectra. ARPES images in a wide energy region before  and after  bismuth deposition are shown in Fig.~\ref{fig1}b and c respectively. Spectrum of pristine MBT surface consists of the set of strongly dispersive  valence band states and parabolic conduction band near the Fermi level. Within the bulk band gap there are topological surface states, but their photoemission cross section strongly depends on the photon energy (see Ref.~[\onlinecite{PhysRevB.100.121104}]), and at the used photon energy of 21.2 eV TSS are difficult to see. Formation of Bi-bilayer on top of MBT results in noticable changes in the electronic spectra.  First, valence band states shift towards the Fermi level by 0.1 eV, similarly to core levels shift, that is caused by the changes of the surface conditions. Moreover, one can see three new hole-like states in the ARPES spectrum, marked as $\alpha, \beta $ and $\gamma$. Two of these states ($\alpha$ and $\beta$) exhibit Rashba-like behaviour and disperse up to 0.3 eV of BE at $\bar \Gamma$-point, while the third one ($\gamma$) approaches the Fermi level. Similar band structure of Bi-bilayer had been observed on various substrates.\cite{Kim_PRB2014, PhysRevB.90.235401, Shen_2019, PhysRevLett.107.166801} 

\begin{figure*}
\includegraphics[width=\textwidth]{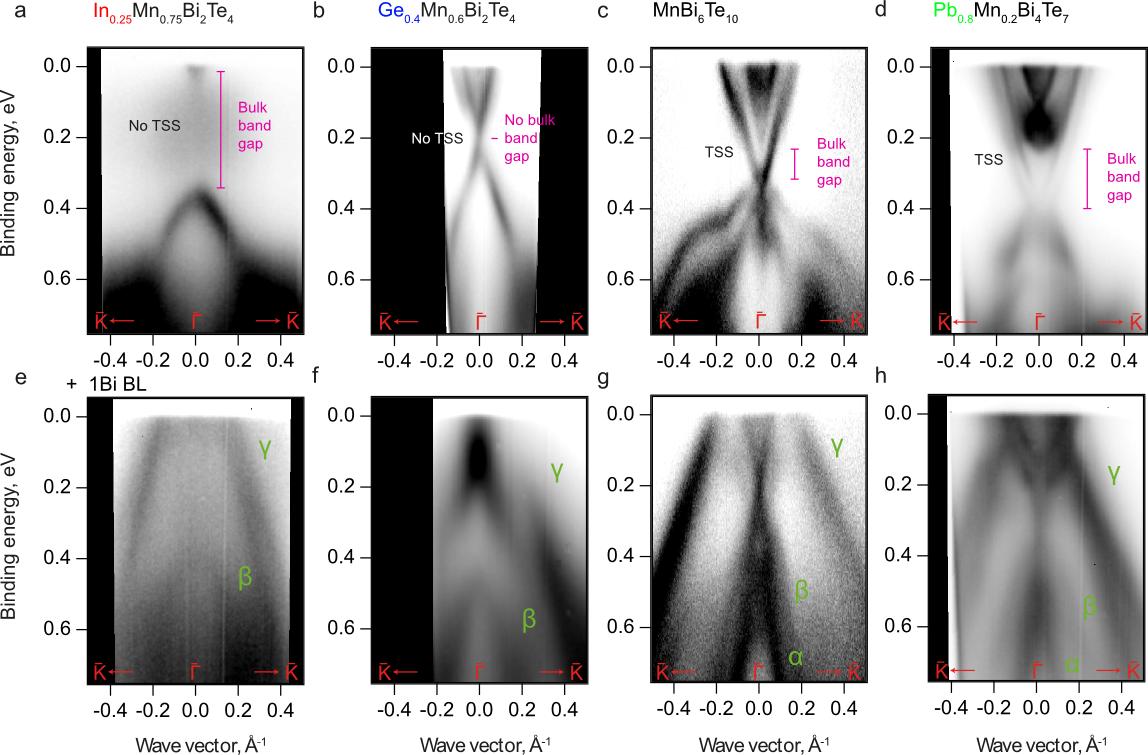}
\caption{
 ARPES dispersion relations in the low energy region in the $\bar{\mathrm{K}}-\bar{\Gamma}-\bar{\mathrm{K}}$ direction of BZ of MnBi$_2$Te$_4$-family samples  taken at  $17\,\mathrm{K}$  before (upper panels) and after (lower panels) Bi deposition. The photon energies are 20 eV for (a,d,e,f,h), 9 eV for (b), 30 eV for (c,g).} 
 \label{fig2}
\end{figure*}

It is known that the electronic structure of free-standing Bi-bilayer is characterized by two degenerate hole-like valence states and electron-like conduction band.\cite{PhysRevLett.107.166801} Giant spin-orbit interaction results in the inversion of the gap between the valence and conduction bands, that turns Bi-bilayer to 2D QSHE topological insulator phase.\cite{Drozdov2014} Growth of Bi-bilayer on the substrate leads to lifting of the inversion symmetry and the appearance of large Rashba-like splitting of the states.\cite{PhysRevLett.107.166801, PhysRevB.89.155116, Kim_PRB2014, PhysRevB.90.235401} Furthermore, Bi-BL and substrate surface states may hybridize,  forming a novel 2D interface electronic system with peculiar topological properties.\cite{Kim_PRB2014, PhysRevB.90.235401, Shen_2019, PhysRevLett.107.166801} Effects of electronic states hybridization in our Bi-BL/MBT system can be analyzed in zoomed region ARPES images, presented in Figs.~\ref{fig1}d-g. Spectrum of pristine MBT surface (Fig.~\ref{fig1}d), taken at a photon energy of 28 eV exhibits the sharp gapped Dirac cone,  formed by topological surface states. The Dirac point gap is of several tens meV, within the limits of previously reported values, ranging from zero to 80-90 meV.\cite{Shikin_2021, Hao_2019}   As noted above, deposition of Bi-bilayer modifies the band structure, introducing several new hole-like Bi-related states, see Fig.~\ref{fig1}e.  In order to resolve this complicated dispersion relations we have calculated  DFT band structure. Calculated spectrum along $\bar\Gamma-\bar{\rm K}$ with projected weights from Bi-BL atoms imposed over the ARPES in Fig.~\ref{fig1}e.  One can see the pronounced Dirac cone-like feature around $\bar \Gamma$-point, interacting  with the hole Bi-related bands. The upper part of the Dirac cone mixes with the band $\gamma$, that is actually split ($\gamma$ and $\gamma$'), with the value of the splitting being dependent on the wave vector, see intensity curvature plot in Fig.~\ref{fig1}g. Similar splitting had been shown for Bi-BL on Bi$_2$Te$_3$\cite{PhysRevLett.107.166801}, and attributed to Rashba-like behaviour due to inversion symmetry breaking.   In the region of lower part of the Dirac cone-like state one can see two hole-like states ($\alpha$ and $\beta$), split by 0.2 eV in energy. The trigonal symmetry of photoemission intensity is seen at the constant energy cuts in Fig.~\ref{fig1}f.  Exchange interaction affects the Bi-BL/MBT interface states, creating an unique QSH/QAH interface, that will be discussed later. 

\subsection*{Bi bilayer on MnBi$_2$Te$_4$-family materials}

Electronic and magnetic structure of MBT-family compounds can be tuned  in two ways: by means of doping and superlattices formation.  Doping  by Pb,Sn or Ge atoms leads to closing of the bulk band gap at some point, forcing the system to become a Dirac semimetal.\cite{frolov2023, estyunina2023, Estyunin2023}  Further doping  leads to reopening of the bulk band gap, but  such a system is not trivial and can be  rather classified as dilute magnetic topological insulator.\cite{frolov2023} On the other hand, exchange interaction and topological properties of MBT  can be varied via superlattice (MnBi$_2$Te$_4$)(Bi$_2$Te$_3$)$_m$ formation, where m=0..6.\cite{Klimovskikh_2020}

We have synthesized Bi bilayer film on top of several MBT-family substrates, namely Ge$_{0.4}$Mn$_{0.6}$Bi$_2$Te$_4$, MnBi$_6$Te$_{10}$ and Pb$_{0.8}$Mn$_{0.2}$Bi$_4$Te$_7$ and novel In$_{0.25}$Mn$_{0.75}$Bi$_2$Te$_4$.   The in-plane lattice constant and surface atomic structure of the compounds are very similar to MBT, and the growth mode of  Bi-BL is also epitaxial, with $1\times 1$ structure. Experimental dispersion relations of the electronic states for pristine surfaces and after Bi-BL adsorption on top  are shown in Fig.~\ref{fig2}.

The first panel (Fig.~\ref{fig2}a) presents the ARPES data for In$_{0.25}$Mn$_{0.75}$Bi$_2$Te$_4$ material, where we see large bulk band gap (of around 350 meV), and no topological surface states within the gap. Thus, one can assume that In doping changes electronic structure of MBT significantly, resulting in trivial insulator phase.   There is some diffuse intensity inside the gap, that can rather be attributed to impurity states or complex scattering processes. Deposition of Bi bilayer on top (Fig.~\ref{fig2}e) leads to appearance of the hole like bands. The first, $\gamma$ one, crosses the Fermi level at $\approx$0.2 \AA$^{-1}$, similar to the $\gamma$ band in Fig.~\ref{fig1} in the Bi-BL/MBT spectrum. One can also see the second hole-like state, that reaches 0.4 eV at $\bar \Gamma$ point and can be related to the $\beta$ state, mixed with valence band of the substrate. Notably, in contrast to Bi-BL on MBT substrate there is no Dirac cone-like feature around the $\bar \Gamma$ point.  

The second panel (Fig.~\ref{fig2}b) shows the spectrum of Ge$_{0.4}$Mn$_{0.6}$Bi$_2$Te$_4$, characterized by a vanishing bulk band gap, making this system a magnetic Dirac semimetal.\cite{frolov2023} One can see the valence and conduction band touching at the BE of around 0.2 eV. Bismuth bilayer deposition on top (Fig.~\ref{fig2}f) of this substrate results in a pair of hole-like bands formation, shifted by 0.1 eV towards the higher BE, in comparison to Bi-BL on MBT and In$_{0.25}$Mn$_{0.75}$Bi$_2$Te$_4$.  Rashba-split $\gamma$ state has large intensity at the $\bar\Gamma$ point where the Kramers degeneracy takes place. Remarkably, the Dirac cone-like features are not visible, similar to Bi-BL on In$_{0.25}$Mn$_{0.75}$Bi$_2$Te$_4$. 

The last two panels of Figure ~\ref{fig2} show the electronic structure of pristine MnBi$_6$Te$_{10}$ (c) and Pb$_{0.8}$Mn$_{0.2}$Bi$_4$Te$_7$ (d) compounds, which are magnetic topological insulators with reduced exchange interaction, in comparison to MBT.\cite{Klimovskikh_2020, Estyunin2023} The crystal structure of these compounds can be viewed as alternating MnBi$_2$Te$_{4}$ septuple layer (SL) (without or with Pb alloying) and Bi$_2$Te$_{3}$ quintuple layer (QL) blocks, and cleaved surface consists of distinct SL or QL  terminations, resulting in multiple Dirac cones structure in ARPES image.  Bismuth bilayer  on top of these two materials exhibits similar features, namely formation of hole-like $\alpha$, $\beta$ and $\gamma$ bands, only with the difference in energy, being by 0.2 eV shifted to higher BE for Pb$_{0.8}$Mn$_{0.2}$Bi$_4$Te$_7$ substrate. Moreover, one can clearly see the Dirac cone feature in both cases, that is  composed of substrate TSS and Bi-BL-related states, similarly to Bi-BL on MBT and Bi$_2$Te$_3$ substrates. Thus, the Bi bilayer exhibits similar set of hole-like bands for all studied MBT-family substrates, but interfacial Dirac cone-like feature is formed only if TSS are present in the bulk band gap of the topological substrate.

\begin{figure*}
\includegraphics[width=\textwidth]{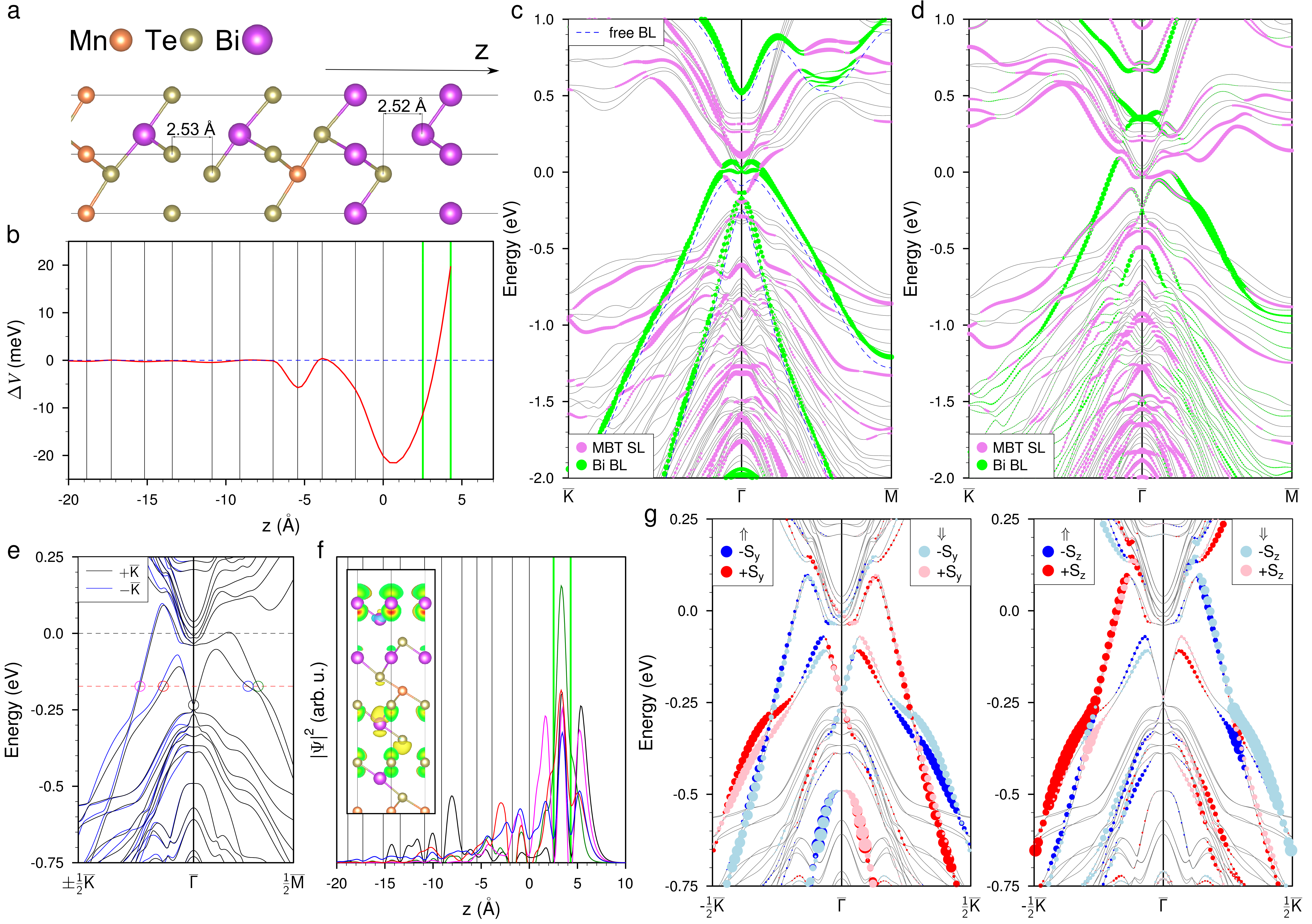}
\caption{
(a) Equilibrium atomic structure of Bi-BL/MBT. (b) 
The interface potential obtained as a difference between the integrated over $xy$ plane $V(z)$ potentials within Bi-BL/MBT heterostructure and free-standing MBT(BL) slab: $\Delta V(z) = V_{\rm Bi-BL/MBT}-V_{\rm MBT(BL)}$. (c) Calculated band structure of Bi-BL/MBT heterostructure when the distance between Bi-BL and MBT surface is twice as large with respect to the equilibrium one. Dashed blue curves show the spectrum of freestanding Bi-BL; size of green and violet dots denote weights of the states localized in the Bi-BL and outermost SL of MBT, respectively. (d) The same as in the panel (c) but for equilibrium position of the Bi-BL. (e) Magnified view of the spectrum shown in panel (d) in the vicinity of the bulk gap of MBT. $\pm\bar{\rm K}$ curves correspond to dispersion of the states along nonequivalent for magnetic MnBi$_2$Te$_4$ $\bar\Gamma - \bar{\rm K}$ and $\bar\Gamma - \text{-}\bar{\rm K}$ directions, respectively. Color circles show states for which the spatial distribution is presented in panel (f) with the same colors. (f) Integrated over $xy$ spatial distributions of the states marked in panel (e). Inset show the charge density of the $\bar\Gamma$ state (black line). (g) Spin texture of the surface states in the Bi-BL/MBT heterostructure along the $\text{-}\bar{\rm K} - \bar\Gamma - \bar{\rm K}$ direction with in-plane ($S_{y}$, left side) and out-of-plane ($S_{z}$, right side) components. Dark and light red/blue circles show positive/negative components for different magnetizaton orientation in the topmost MnBi$_2$Te$_4$ SL: along ($\Uparrow$) and against ($\Downarrow$) the surface normal, respectively.}
 \label{fig3}
\end{figure*}

\subsection*{DFT calculations}

To clarify the interaction of the deposited Bi-BL with MBT substrate and reveal how the formation of the interface affects the electronic structure of the Bi-BL we performed DFT calculations.
In the recently proposed vdW heterostructure composed of single MnBi$_2$Te$_4$ SL and Bi-BL \cite{Xue_arxiv_2212_12905} the calculated magnetic easy axis is predicted to be in-plane, with a magnetic anisotropy energy (MAE) of $E_x-E_z = -0.15$ meV/Mn that is different from the out-of-plane magnetic easy axis of the free-standing MnBi$_2$Te$_4$ SL where MAE is $+0.125$ meV/Mn \cite{Otrokov.prl2019}. For our structure with thick MnBi$_2$Te$_4$ substrate we check the possibility to change the spin alignment in the outermost MnBi$_2$Te$_4$ SL due to Bi-BL interface formation. We found that total energy of the structure where magnetic moments of Mn atoms in the interfacial SL are placed in-plane is 0.07 meV/Mn less favorable compared to the case when the intrinsic out-of-plane magnetization of MnBi$_2$Te$_4$ is preserved, although this value is smaller than MAE in the free-standing SL.

\subsubsection*{Bi-BL/MBT surface electronic structure}

The equilibrium structure of Bi-BL on the MBT surface is shown in Fig.~\ref{fig3}a. 
The optimized interface distance between Bi-BL and topmost Te atomic layer of the MBT substrate is 2.52 \AA, which is comparable to the vdW spacings in MBT bulk (2.53 \AA) \cite{Eremeev.jac2017,Otrokov_2019_Nature}. Our calculations for interactions between Bi and Te atoms at the Bi-BL/MBT interface and those (Te-Te) at the vdW spacing in the MBT, based on the projected crystal orbital Hamilton population method \cite{Dronskowski1993,Deringer2011,LOBSTER-2016} show that the former one is only by factor 1.6 stronger (0.49 vs. 0.30 eV). Thus, the interaction between MnBi$_2$Te$_4$ surface and Bi-BL is close to the vdW type. At that, the charge transfer from Bi-BL to the substrate is negligibly small: it is only 0.14 $e$ from the interfacial Bi layer and completely absent in the upper Bi layer. Such a small charge transfer is reflected in minor variation in the interface potential: $\Delta V(z)$ obtained as a difference between the electrostatic potentials within Bi-BL/MBT heterostructure and free-standing MBT(BL) does not exceed 20 meV in both MBT surface and Bi-BL film (Fig.~\ref{fig3}b).

When the distance between Bi-BL and the surface is enlarged twice, up to 5 \AA, Fig.~\ref{fig3}c, one can clearly see the Bi-BL bands which lie a bit higher than in the spectrum of the free-standing Bi-BL (blue dashed lines). These bands demonstrate moderate Rashba-type splitting due to the inversion symmetry breaking. The gapped topological surface state of the MBT surface is also present in the spectrum. Despite the practically absent interaction between the bilayer and the surface in this case the TSS hybridizes with upper and second valence bands of the Bi-BL with the formation of small hybridization gaps at the Fermi level and at $\approx -0.15$ eV, respectively. In contrast, at equilibrium interface distance (Fig.~\ref{fig3}d) the spectrum of the heterostructure undergoes significant changes, becoming strongly $\bar\Gamma-\bar{\rm K}$/$\bar\Gamma-\bar{\rm M}$ anisotropic and  demonstrating mixed BL/SL localization of the surface states within MBT bulk gap. Such a behavior is typical for the TSSs which tend to penetrate into the overlayer \cite{Menshov_JPCM2014}. This effect is observed on both nonmagnetic \cite{Eremeev_SciRep2015,DeLuca_PRM2023} and magnetic \cite{Zaitsev_PRB2023} TIs and also underlies the formation of magnetic extension heterostructures \cite{Hirahara,Otrokov.jetpl2017,Otrokov.2dmat2017}. As can be seen in Figs.~\ref{fig3}e,f all surface states in the Bi-BL/MBT heterostructure locating mostly in the Bi-BL overlayer penetrate deep into the MBT substrate down to the second SL. Owing to their predominant localization in the Bi bilayer, these states generally inherit the dispersion of the Bi-BL and are similar to the states in the heterostructure formed by Bi-BL on a nonmagnetic TI \cite{Kim_PRB2014}, but due to their penetration into the magnetic substrate, the  $\bar\Gamma$ degeneracies are lifted. 
Another effect of magnetic substrate on the surface states arising from hybridization of the Dirac state of MBT and Rashba state of Bi-BL is appearance of $\bar\Gamma-\pm\bar{\rm K}$ asymmetry 
(Fig.~\ref{fig3}e), 
similar to that in the Dirac state of MnBi$_2$Te$_4$ \cite{Otrokov_2019_Nature}. This asymmetry, is absent along $\bar\Gamma-\bar{\rm M}$ where the mirror plane symmetry obliges the out-of-plane spin components to be zero for both Rashba and Dirac states. However, for the $\bar\Gamma-\bar{\rm K}$ directions the nonzero $S_z$ components are allowed by symmetry and they coexist with the in-plane spin components. Since the surface states do not penetrate deeper than the topmost SL, where Mn magnetic moments are ferromagnetically ordered, the $\bar\Gamma-\text{+}\bar{\rm K}$ and $\bar\Gamma-\text{-}\bar{\rm K}$  branches with nonzero $S_z$ interact differently with the Zeeman field provided by the Mn layer of the topmost SL block and this interaction and resulting alteration in the $\bar\Gamma-\pm\bar{\rm K}$ dispersions depends on the magnetization direction in the topmost SL (Fig.~\ref{fig3}g).

\begin{figure*}
\includegraphics[width=\textwidth]
{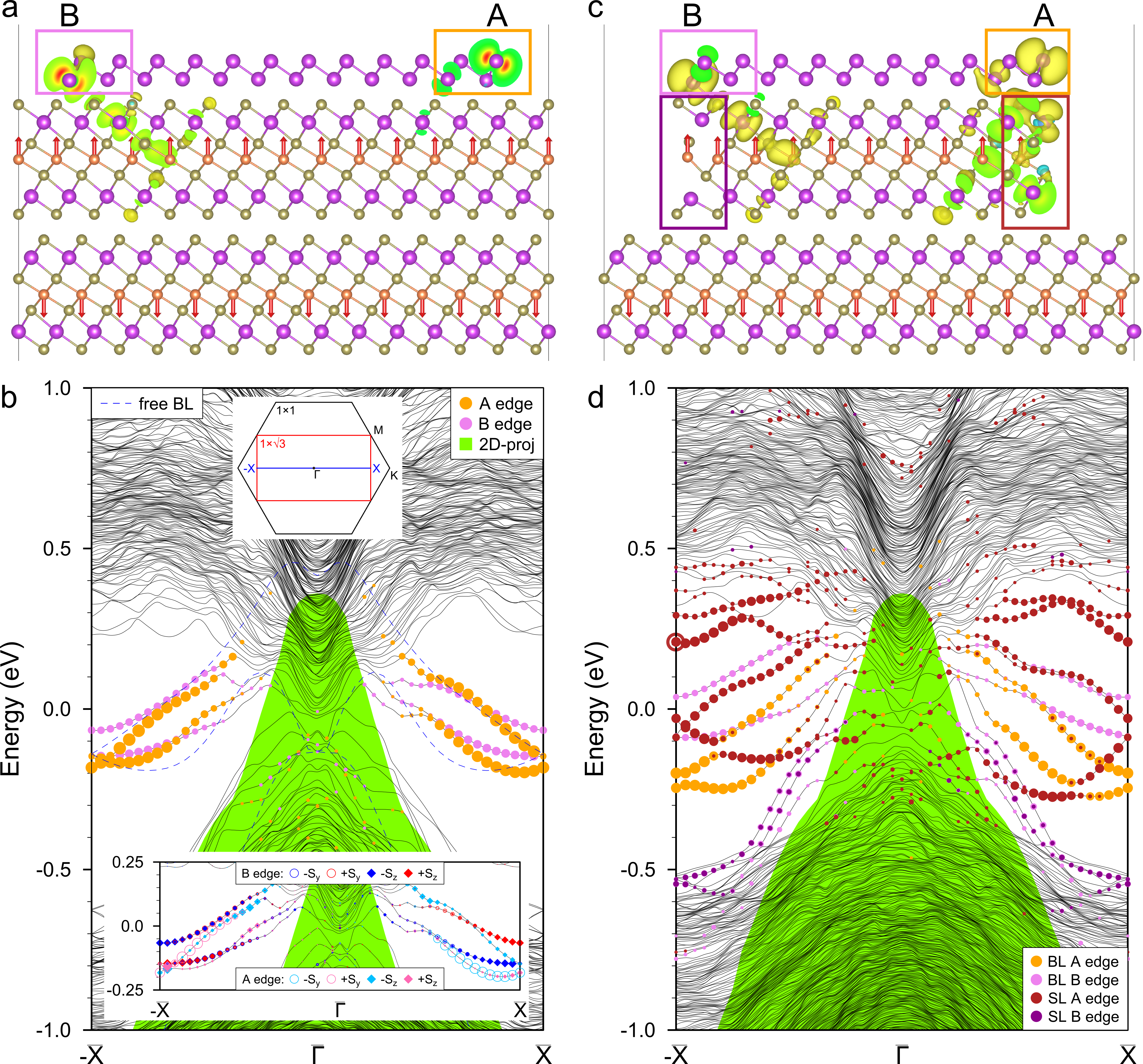}
%{Fig4.png}
%\includegraphics[width=\textwidth]{Fig4_v2.pdf}
\caption{
(a) Side view of the atomic structure of $1\times 7\sqrt{3}$ MnBi$_2$Te$_4$ surface supercell with Bi-BL ribbon ontop and spacial charge distribution of the 1D states localized on right (A-type) and left (B-type) ribbon edges. Colored boxes show the areas over which the atomic weights of the states (shown in panel (b) in the same colors) were collected. Red arrows show Mn magnetic moments. (b) Edge band structure of Bi-BL ribbon on flat MnBi$_2$Te$_4$ surface calculated along $\text{-}\bar{\rm X}-\bar\Gamma-\bar{\rm X}$ direction of the 1D BZ (see top inset). Green area depicts projection of 2D states (hybridized Bi-BL/MBT states), crossing the MBT bulk gap, onto 1D BZ. Orange and violet circles show weights of the 1D states localized on A and B edges of the ribbon, respectively. Dashed blue curves show the spectrum of zig-zag edge states in the freestanding Bi-BL ribbon. Bottom inset demonstrate spin texture of the edge states.  (c and d) The same as in panels (a,b) but in this case the Bi-BL ribbon %is not located on the flat terrace, but 
is connected with a step on the MBT surface. Weights of the states localized on Bi-BL and MBT-SL edges are shown separately with colors pointed in the key. }
 \label{fig4}
\end{figure*}

\subsubsection*{Bi-BL edge electronic structure on flat and stepped MBT surfaces}

Further, we have calculated the electronic structure of  Bi-bilayer nanoribbon on top of MBT, in order to analyze whether Quantum Spin Hall phase of Bi-BL survives or not. For this end we constructed the $1\times 7\sqrt{3}$ MBT supercell with zig-zag edged Bi-BL ribbon on the top surface (Fig.~\ref{fig4}a). In this supercell in the AFM MBT slab the Mn magnetic moments in the topmost SL were pointed outward the surface plane. The calculated edge spectrum of the freestanding Bi-BL ribbon (Fig.~\ref{fig4}b, dashed blue curves) agrees well with earlier work \cite{Wada_PRB2011}.  
Note that the calculations of Bi-BL ribbon on nonmagnetic TI \cite{Kim_PRB2014} revealed that overall band dispersion, the Dirac cone at the zone boundary and the helical spin texture of the edge states are preserved and consistent qualitatively with that of a free-standing Bi(111) BL despite of strong hybridization of the Bi-BL bands with the substrate. As can be seen in the Fig.~\ref{fig4}b the dispersion of the topological edge states of Bi-BL on MBT AFMTI also survives far from the BZ center in which the states fall into 2D continuum of the hybridized Bi-BL/MBT states crossing the MBT bulk gap (green area in Fig.~\ref{fig4}b). The difference from the spectrum of free-standing bilayer is that the spectra of left and right zig-zag edges, which were degenerate in the free BL, are different on the MBT substrate due to different atomic relaxation of the opposite edges and different spatial localization of the edge states. Note that the similar behavior is observed for the Bi bilayer on other substrates and the edges are usually called as A- and B-type edges \cite{Yeom_PRB2016} (Fig.~\ref{fig4}a). The state localized on the left (B-type) edge having maximum charge localization on the most extreme atoms of the BL ribbon rapidly penetrates deep into the topmost SL while on the opposite (A-type) edge the 1D state propagates within the BL and only partially penetrates into the substrate. 
This penetration of the edge states into magnetic substrate results in opening of the gaps in the Dirac states at the BZ boundaries. The exchange 
gap in Dirac point of the A-type  edge state amounts 36 meV while it is much larger, of 82 meV, for the B-type  edge state, which penetrates well to the manganese layer. Similarly to $\bar\Gamma -\pm \bar{\rm K}$ asymmetry in the 2D spectrum of the Bi-BL/MBT the Bi-BL edge spectrum also demonstrate $\bar\Gamma -\pm \bar{\rm X}$ asymmetry (note that 1D BZ direction coincides with the $\text{-}\bar{\rm K}-\bar\Gamma - \bar{\rm K}$ direction the the 2D BZ, see top inset in Fig.~\ref{fig4}b). This asymmetry stem from interaction of the spin-polarized edge states demonstrating considerable $S_z$ spin component (bottom inset in Fig.~\ref{fig4}b) and Zeeman field of the topmost SL. It is evident that the presented in Fig.~\ref{fig4}b dispersion of the edge states along $\text{-}\bar{\rm X}-\bar\Gamma - \bar{\rm X}$ will be reversed when the direction of the Mn magnetic moments in the topmost SL will be opposite.

It is known the MBT surfaces usually contain steps of SL  height and change in the direction of the surface layer magnetization in MBT AFMTI is related to these steps\cite{Sass2020}. Despite the  terraces are usually wide and step density is relatively small we also consider the Bi bilayer edge coinciding with the step edge on the MBT substrate in order to find out whether the edge states of the Bi bilayer survive in such a geometry. Besides, in the recently proposed free-standing vdW heterostructure composed of single MnBi$_2$Te$_4$ SL and Bi-BL \cite{Xue_arxiv_2212_12905} in case of out-of-plane magnetization within SL the TB calculations predict the emergence of topologically protected edge states. In our model the constructed ribbon composed of Bi-BL and MBT SL (Fig.~\ref{fig4}c) presents such a heterostructure, however, forming in a natural way at the edge of the step and supported by MBT bulk. The results show (Fig.~\ref{fig4}d) that MBT gap region possess a plenty of the states originating from the dangling bonds of the SL edges (the spatial localization only one of them, marked with open brown circle in the panel d is shown in the panel c). In addition to these states in the spectrum, one can easily find out the states of the Bi-BL edges (violet and orange circles in panel d). In this case, the dispersion of A-type  edge state on the whole is shifted lower in energy compared to the Bi-BL on the flat terrace while that of the B-type  edge is shifted up so that the upper branch becomes completely unoccupied. Besides, the $\pm \bar{\mathrm{X}}$  Dirac point exchange gaps are enlarged for both B-type  and A-type  edge states, up to 122 and 46 meV, respectively, which should indicate an increased  overlap of the edge states with the manganese layer, although in general the spatial distribution of the edge states (Fig.~\ref{fig4}c) looks similar to that on the flat terrace. It is also worth noting that the $\bar\Gamma -\pm \bar{\rm X}$ asymmetry, controlled by the direction of magnetization in the upper Mn layer is preserved in this case as well. 

Comparing our spectrum containing states resided on Bi-BL edges and numerous states generated by the SL edges with that for free-standing BL-SL heterostructure with the same magnetization obtained within TB calculation \cite{Xue_arxiv_2212_12905} without edge potential taken into account, we cannot confidently relate which of these states could be the  earlier predicted topological states of the BL-SL heterostructure, and besides, they are all gapped in our case.

It should also be noted that STM topographic measurements of MnBi$_2$Te$_4$ surface show that the edges of the SL steps have a more complex and variable geometry \cite{Garnica.npj2022}  than the edges used in our simulation, which are strictly perpendicular to the surface, and hence the dispersion of the states localized on the SL edges in real samples will differ from obtained in the calculation.

In summary,  we have fabricated and studied a number of heterostructures composed of Bi bilayer and MnBi$_2$Te$_4$-family materials. Similar lattice constants of Bi(111) and IMTI compounds results in epitaxial growth mode and $1 \times 1$ structure  of the Bi film. Electronic states of Bi-BL and IMTI hybridize, forming a joint interface band structure that is revealed by means of ARPES and DFT. Bismuth states mix with the topological surface states and bulk bands of IMTI dependently on the substrate composition, that is expected to affect the induced exchange interaction. By means of $ab~initio$ calculations we have shown how the exchange interaction produces splitting of 1D edge states in the case of Bi nanoribbon, that makes this heterostructure promising for tunable and yet unobserved quantum Hall phases detection. 

\noindent

\section*{Methods}

Intrinsic magnetic topological MnBi$_2$Te$_4$ compounds had been grown by Bridgeman method.  Characterization of the MnBi$_2$Te$_4$-family monocrystals can be found in Refs.~[\onlinecite{Klimovskikh_2020, Otrokov_2019_Nature, Shikin.prb2021, frolov2023}]. ARPES data were recorded using several facilities. Laboratory-based measurements were
made using a SPECS GmbH ProvenX-ARPES system located in ISP SB RAS equipped with ASTRAIOS 190 electron energy analyzer with 2D-CMOS electron detector and a non-monochromated He I$\alpha$ light source with h$\nu$=21.22 eV. ARPES spectra at various photon energies were measured at the BaDElPh beamline\cite{PETACCIA2009780} of the Elettra synchrotron in Trieste and LOREA beamline of the ALBA synchrotron in Barcelona. All measurements were carried out at T = 17 K using p-polarized photons.   

Samples were cleaved \emph{in situ} at the base pressure of  $6\times 10^{-10}\,\mathrm{mbar}$.  Crystal quality and surface cleanliness  of the cleaved surfaces had been checked by low energy electron diffraction and XPS. Bismuth deposition has been done using standart Knudsen cell at the temperature of 700 K. The sample was held at room temperature during deposition, and had been annealed up to 500 K after film growth.  Bi deposition rate has been checked by quarz microbalance. Additionally, the thickness of 1 BL of Bi has been verified by the procedure of XPS spectra fitting and analysis, ascribed in the manuscript, and double checked by means of comparison with Bi deposition on Bi$_2$Te$_3$ substrate.

Electronic structure calculations were carried out within the density functional theory using the projector augmented-wave (PAW) method \cite{Blochl.prb1994} as implemented in the VASP code \cite{vasp1,vasp2}. The exchange-correlation energy was treated using the generalized gradient approximation \cite{Perdew.prl1996}. The Hamiltonian contained scalar relativistic corrections and the spin-orbit coupling was taken into account by the second variation method \cite{Koelling.jpc1977}. 
The crystal structure of MnBi$_2$Te$_4$ bulk was fully optimized to find the equilibrium lattice parameters and atomic positions. At that, a conjugate-gradient algorithm was used. In order to describe the van der Waals interactions we made use of the DFT-D3 approach with Becke-Johnson damping \cite{Grimme2011}. Spin-orbit coupling was always included when performing relaxations.
The atomic coordinates were relaxed using a force tolerance criterion for convergence of 10$^{-4}$ eV/{\AA} and the convergence criterion for the total energy was 10$^{-6}$ eV.
The $k$-point meshes of 10$\times10\times$2 and 10$\times10\times$1 were used to sample the the bulk and slab Brillouin zones, respectively.
The Mn $3d$-states were treated employing the GGA$+U$ approach \cite{Anisimov1991} within the Dudarev scheme \cite{Dudarev.prb1998}. The $U_\text{eff}=U-J$ value for the Mn 3$d$-states was chosen to be equal to 5.34~eV, as in previous works on MnBi$_2$Te$_4$ \cite{Otrokov.jetpl2017,Otrokov.2dmat2017,Eremeev.jac2017,Eremeev.nl2018, Otrokov_2019_Nature,Otrokov.prl2019, Klimovskikh_2020,Hirahara2020,Shikin.srep2020,Shikin.prb2021, Garnica.npj2022}.

To simulate the Bi-bilayer on MnBi$_2$Te$_4$ surface a slab consisting of 6 septuple layers of MnBi$_2$Te$_4$ with optimized bulk lattice constant was used. The atomic positions of two Bi bilayers atoms (on the top and bottom sides of the slab) as well as topmost(lowest) SLs of the MnBi$_2$Te$_4$ slab were optimized. Substrate atoms of the central SLs were kept fixed at the bulk crystalline positions. The geometry optimization was performed until the residual force on atoms was smaller than 10 meV/\AA.

To simulate the zig-zag edge of the Bi-bilayer we constructed the rectangular $1\times 7\sqrt{3}$ MnBi$_2$Te$_4$ supercell the thickness of which was reduced to four SLs. Ontop of this supercell a Bi-bilayer ribbon of $\approx$41 \AA\ width was placed. The distance between opposite edges of the ribbon in neighboring cells was about 10 \AA. The atoms on the ribbon edges were relaxed up to the fifth atom from the edge. 

\section*{Acknowledgements}

I.I.K and T.V. acknowledge the support from the Red guipuzcoana de Ciencia, Tecnología e Innovación – Gipuzkoa NEXT 2023 from the Gipuzkoa Provincial Council. The authors acknowledge support by Russian Science Foundation grant № 22-72-10074 and Saint Petersburg State University (Grant No. ID 95442847). S.V.E. acknowledge support from the Government research assignment for ISPMS SB RAS, project FWRW-2022-0001 (in the part of the DFT calculations). 
The authors also acknowledge support by Russian Foundation for Basic Research (Grants No. 21-52-12024, and No. 19-29-12061) and state assignment of IGM SB RAS 122041400031-2 and  ISP SB RAS. The calculations were partially performed using the equipment of the Shared Resource Center “Far Eastern Computing Resource” of IACP FEB RAS (https://cc.dvo.ru).
S.B-C thanks the MINECO of Spain, project PID2021-122609NB-C21. LOREA was co-funded by the European Regional Development Fund (ERDF) within the Framework of the Smart Growth Operative Programme 2014-2020. We acknowledge Jordi Prat for technical support during ARPES experiments at LOREA.
\bibliography{aaa.bib}

\end{document}